\documentclass[%
superscriptaddress,
showpacs,preprintnumbers,
nofootinbib,
amsmath,amssymb,
aps,
prd,
tightenlines,
11pt,
notitlepage
]{revtex4-1}
\usepackage[T1]{fontenc}

\usepackage{alltt}
\usepackage{amsmath}
\usepackage{amsthm}
\usepackage{amssymb}
\usepackage{bm}
\usepackage{booktabs,siunitx}
\usepackage{cancel}
\usepackage{color}
\usepackage{csquotes}
\usepackage{enumerate}
\usepackage{graphicx,bm}
\usepackage{lipsum}
\usepackage{listings}
\usepackage{mathtools}
\usepackage{mathrsfs}
\usepackage{physics}
\usepackage{relsize}
\usepackage{slashed}
\usepackage{soul}
\usepackage{spverbatim}
\usepackage[caption=false]{subfig}
\usepackage{wrapfig}
\usepackage{xcolor}
\usepackage{diagbox}
\usepackage{multirow}

\usepackage[utf8]{inputenc}

\usepackage[pdftex,hypertexnames=false,linktocpage=true]{hyperref}
\hypersetup{colorlinks=true,linkcolor=blue,anchorcolor=blue,citecolor=darkgreen,filecolor=blue,urlcolor=blue,bookmarksnumbered=true,
	pdfview=FitB
}
\colorlet{darkgreen}{green!60!black}
\colorlet{brightyellow}{yellow!75!red}
\colorlet{orange}{red!50!yellow}
\colorlet{darkblue}{blue!60!black}
\colorlet{darkred}{red!80!black}
\colorlet{greenblue}{green!50!blue}

\makeatletter

\newcommand{\Rmnum}[1]{\expandafter\@slowromancap\romannumeral #1@}
\makeatother

\def\dd{{\mathrm{d}}}
\def\imag{{\mathrm{i}}}

\begin{document}
	
	\title{$B_c$ mesons and their properties on the light front}
	
	\author{Shuo Tang}
	\email{tang@iastate.edu}	
	\affiliation{%
		Department of Physics and Astronomy, Iowa State University, Ames, IA 50011, USA
	}%
	
	\author{Yang Li}
	\affiliation{%
		Department of Physics, College of William \& Mary, Williamsburg, VA 23185, USA
	}	
	
	\author{Pieter Maris}
	\affiliation{%
		Department of Physics and Astronomy, Iowa State University, Ames, IA 50011, USA
	}
	
	\author{James P. Vary}
	\affiliation{%
		Department of Physics and Astronomy, Iowa State University, Ames, IA 50011, USA
	}
	
	
	
	
	\begin{abstract}
		We investigate the unequal mass relativistic bound state system, the $B_c$ mesons, in a light-front Hamiltonian formalism. We adopt an  effective Hamiltonian based on soft-wall light-front holography, together with a longitudinal confinement that was first introduced for heavy quarkonium. We also include the one gluon-exchange interaction with a running coupling, which produces the spin structure. We present the mass spectrum and light-front wave functions. We also use the light-front wave functions to calculate decay constants, charge and momentum densities, as well as distribution amplitudes. The results are compared with experiments and other theoretical approaches, all of which are in reasonable agreement.
	\end{abstract}
	\maketitle
	
	\section{Introduction}
	
	The $B_c$ system is the only known heavy meson family with unequal quark masses, which provides an important testing ground for understanding strong interaction physics. 
	While the spectra and properties of the $c\bar c$ and $b\bar b$ mesons are extensively studied in experiments, data on $b\bar c$ or $c \bar b$ are relatively scarce.
	Until now, only two states, the ground state and its first radial excitation, are confirmed in experiments \cite{PhysRevD.58.112004,PhysRevLett.113.212004}. 
	Meanwhile, ongoing and forthcoming high energy experiments, e.g. LHC and RHIC, are expected to generate a large ensemble of these particles. For these reasons, there are renewed interests in theoretical investigations \cite{PhysRevD.96.054501,PhysRevD.91.114509,PhysRevD.94.034036,BHATTACHARYA2017430}. 
	
	Light-front quantization is a natural relativistic framework to describe the intrinsic partonic structure of hadrons  \cite{BAKKER2014165}. Among various light-front approaches, light-front holographic (LFH) models stand out as semi-classical approximations to QCD (see Ref. \cite{BRODSKY20151} and the references therein).  Meanwhile, a computational framework known as the basis light-front quantization (BLFQ), 
	has been established to tackle the many-body dynamics and has been applied to QED \cite{maris2013bound,PhysRevD.91.105009} and QCD \cite{LI2016118,PhysRevD.96.016022} bound states.  In the latter case, LFH is embedded in the BLFQ formulation to model the heavy quarkonium (charmonium and bottomonium) system. The results have shown good agreement with experiments and other theoretical models. 
	
	In this work, we adapt the successful Hamiltonian of Refs. \cite{LI2016118,PhysRevD.96.016022} to the $B_c$ system in the BLFQ approach. In essence, this model implements the AdS/QCD soft-wall Hamiltonian \cite{PhysRevLett.102.081601} plus a longitudinal confinement \cite{LI2016118}, both of which are of long range. In addition, we adopt the one-gluon exchange with a running coupling \cite{PhysRevD.96.016022}. This term controls the short-distance physics and embeds the spin structure information. We solve the $B_c$ system without introducing any additional free parameter (other than the ones employed in charmonium and bottomonium). Therefore, it is also a test of the predictive power of the model proposed in Ref. \cite{PhysRevD.96.016022}. This work is a straightforward yet necessary step for developing a relativistic model for hadrons based on light-front holography and light-front dynamics.
	
	We begin by introducing the effective light-front Hamiltonian and the basis function approach in Sec. \ref{sec2}, following Ref. \cite{PhysRevD.96.016022}. Presented in Sec. \ref{sec3} are results including the mass spectrum, wave functions, decay constants, transverse charge and momentum density, and distribution amplitudes. They are compared with experiments and other theories whenever available. We also discuss the differences between $B_c$ and heavy quarkonium. Sec. \ref{sec4} summarizes our current work and provides a brief discussion of possible improvements.
	
	\section{\label{sec2}Hamiltonian Formalism and the Basis Function Representation}
	
	The light-front Hamiltonian formalism leads to an eigenvalue equation $H\ket{\psi_h} = M^2_h\ket{\psi_h} $. Here we adapt the effective Hamiltonian of Refs. \cite{LI2016118,PhysRevD.96.016022,PhysRevLett.102.081601} for unequal quark masses:
	\begin{equation}
	\begin{split}
	H_\text{eff} = \frac{\vec{k}^2_\bot+m^2_q}{x} + \frac{\vec{k}^2_\bot+m^2_{\bar{q}}}{1-x} +\kappa^4 \vec{\zeta}^2_\bot  - \frac{\kappa^4}{(m_q+m_{\bar{q}})^2}&\partial_x\big(x(1-x)\partial_x\big) \\
	-&\frac{C_F 4\pi \alpha_s(Q^2)}{Q^2}\bar{u}_{s'}(k')\gamma_\mu u_s(k) \bar{v}_{\bar{s}}(\bar{k})\gamma^\mu v_{\bar{s}'}
	(\bar{k}'),
	\end{split}
	\end{equation}
	where $\vec{\zeta}_\bot \equiv \sqrt{x(1-x)} \vec{r}_\bot$ is the holographic variable \cite{BRODSKY20151}, $C_F = (N_c^2-1)/(2N_c)=4/3$ is the color factor of the $q\bar{q}$ color singlet state. In this paper, we investigate the $B_c$ system as $b\bar{c}$, i.e. $B_c^-$. Therefore $m_q$ is the mass of the bottom quark and $m_{\bar{q}}$ the anti-charm quark. $x$ and $(1-x)$ are the longitudinal momentum fractions of $b$ and $\bar{c}$, respectively.
	We incorporate a running coupling for the one-gluon exchange potential, which is modeled as  \cite{PhysRevD.96.016022}, 
	\begin{equation}
	\alpha_s(Q^2)=1/[ \beta_0 \ln (Q^2/\Lambda^2+\tau) ],
	\end{equation}
	where $\beta_0=(33-2N_f)/(12\pi)$, with the quark flavor number taken to be $N_f=4$. We use $\Lambda = 0.13$ GeV and in order to avoid the pQCD IR divergence we use $\tau = 12.3$ such that  $\alpha(0) = 0.6$. See Ref. \cite{PhysRevD.96.016022} for more details.
	
	
	
	We adopt a Fock space limited to the $\ket{q\bar{q}}$ sector where the state vector reads,
	\begin{equation}
	\begin{split}
	\ket{\psi_h(P,j,m_j)}=\sum_{s,\bar{s}}\int_0^1&\frac{\dd x}{2x(1-x)}\int\frac{\dd ^2 k_\bot}{(2\pi)^3}\ \psi_{s\bar{s}/h}^{(m_j)}(\vec{k}_\bot,x)\\
	&\times\frac{1}{\sqrt{N_c}}\sum_{i=1}^{N_c} b^\dagger_{si/b} (xP^+,\vec{k}_\bot+x\vec{P}_\bot) d^\dagger_{\bar{s}i/\bar{c}} ((1-x)P^+,-\vec{k}_\bot+(1-x)\vec{P}_\bot)\ket{0}.
	\end{split}
	\end{equation}
	In the expression above, $\psi_{s\bar{s}/h}^{(m_j)}(\vec{k}_\bot,x)$ represents the light-front wave functions (LFWFs), $s$ and $\bar{s}$ are the spins of the quark and anti-quark, respectively.
	The anti-commutation relations of creation operators and the orthonormal relation of state vectors in this work are similar as for heavy quarkonium \cite{PhysRevD.96.016022}.
	
	
	We use a  basis function approach, BLFQ \cite{PhysRevC.81.035205}, and following Ref. \cite{LI2016118}, we represent the LFWFs in terms of transverse and longitudinal basis functions $\phi_{nm}$ and $\chi_l$, with basis coefficients $\psi_h(n,m,l,s,\bar{s})$,
	\begin{equation}
	\psi_{s\bar{s}/h}^{(m_j)}(\vec{k}_\bot,x) =\sum_{n,m,l} \psi_h(n,m,l,s,\bar{s})\phi_{nm}(\vec{k}_\bot/\sqrt{x(1-x)})\chi_l(x).
	\end{equation}
	For the basis functions, we employ
	\begin{equation}
	\begin{gathered}
	\phi_{nm}(\vec{q}_\bot;b)=\frac{1}{b}\sqrt{\frac{4\pi n!}{(n+|m|)!}} \bigg(\frac{q_\bot}{b}\bigg)^{|m|} e^{-\frac{1}{2}q^2_\bot/b^2} L^{|m|}_n (q^2_\bot/b^2) e^{\imag m\theta_q} ,\\
	\chi_l(x;\alpha,\beta)=\sqrt{4\pi(2l+\alpha+\beta+1)}\sqrt{\frac{\Gamma(l+1)\Gamma(l+\alpha+\beta+1)}{\Gamma(l+\alpha+1)\Gamma(l+\beta+1)}} x^{\frac{\beta}{2}}(1-x)^{\frac{\alpha}{2}} P^{(\alpha,\beta)}_l(2x-1),
	\end{gathered}
	\end{equation} 
	which are the analytical solutions of the effective Hamiltonian without the one-gluon exchange. 
	Here, $\phi_{nm}$ is the 2D harmonic oscillator function with $n$ and $m$ the principle and orbital quantum numbers, respectively, with $\vec{q}_\bot = \vec{k}_\bot/\sqrt{x(1-x)}, \ q_\bot= \abs{\vec{q}_\bot}, \ \theta_q=\arg \vec{q}_\bot$, and $L^{\abs{m}}_n(z)$ is the associated Laguerre polynomial. Note that the conserved total magnetic projection $m_j$ is the sum of the orbital projection $m$, and the sum of the spin projections, $m_j = m + s + \bar{s}$. We adopt $b=\kappa$ for the scale parameter in the HO basis. For the longitudinal basis function $\chi_l$, $l$ is the longitudinal quantum number, $P^{(\alpha,\beta)}_l(z)$ is the Jacobi polynomial . The dimensionless parameters $\alpha$ and $\beta$ are associated with the quark masses: $\alpha = 2m_{\bar{c}}(m_b+m_{\bar{c}})/\kappa^2$ and $\beta = 2m_b(m_b+m_{\bar{c}})/\kappa^2$. 
	%
	When the one-gluon exchange is implemented, one can solve the eigen-equation by diagonalizing the Hamiltonian matrix. Hence the obtained eigenvalues represent the spectra as squared masses, and the eigenvectors are the coefficients $\psi_h(n,m,l,s,\bar{s})$.

	\section{\label{sec3}Numerical results}
	\begin{table} \footnotesize
		\centering 
		\begin{tabular}{ccccccccccc}
			\hline\hline
			&  \hspace{0.1cm} $N_f$  \hspace{0.1cm} &   \hspace{0.1cm}$\kappa$ (GeV)   \hspace{0.1cm}&  \hspace{0.1cm} $m_c$ (GeV)  \hspace{0.1cm} &  \hspace{0.1cm} $m_b$ (GeV)  \hspace{0.1cm}  &  \hspace{0.1cm}  r.m.s (MeV)   \hspace{0.1cm}&  \hspace{0.1cm} $\overline{\delta_j M}$ (MeV)  \hspace{0.1cm}&  $N_\text{max} = L_\text{max}$  &  Ref. \\
			\hline
			$c\bar{c}$ &$4$& $0.966$ & $1.603$&$-$&$31$&$17$&$32$&\cite{PhysRevD.96.016022} \\
			$b\bar{b}$  &$5$& $1.389$ &$-$&$4.902$&$38$&$8$&$32$&\cite{PhysRevD.96.016022} \\
			$b\bar{c}$  &$4$& $1.196$ &$1.603$&$4.902$&$37$&$6$&$32$&this work	\\
			\hline\hline
		\end{tabular}
		\caption{Summary of the model parameters}
		\label{tb1}
	\end{table}
	
	In this work, we adopt model parameters from those of the charmonium and bottomonium calculations without doing further parameter fitting. In particular, we adopt the quark masses from the charmonium and bottomonium applications \cite{PhysRevD.96.016022} (See TABLE \ref{tb1}). On the other hand, the confining strength is taken as $\kappa_{b\bar{c}} = \sqrt{(\kappa^2_{c\bar{c}}+\kappa^2_{b\bar{b}})/2}$, where $\kappa_{b\bar{b}}$ and $\kappa_{c\bar{c}}$ are the confining strength of the charmonium and bottomonium system, respectively. This is in accordance with the heavy quark effective theory (HQET) \cite{PhysRevD.95.034016}. All the calculations in this paper, unless otherwise stated, are based on $N_\text{max} = L_\text{max}=32$, which is associated with UV and IR regulators $\Lambda_\text{UV} = b\sqrt{N_\text{max}} \simeq 6.77$ GeV  and $\lambda_\text{IR} = b / \sqrt{N_\text{max}}\simeq 0.21$ GeV.

	\subsection{Mass Spectroscopy}
	
\begin{figure}
		\centering
		\includegraphics[scale=0.45]{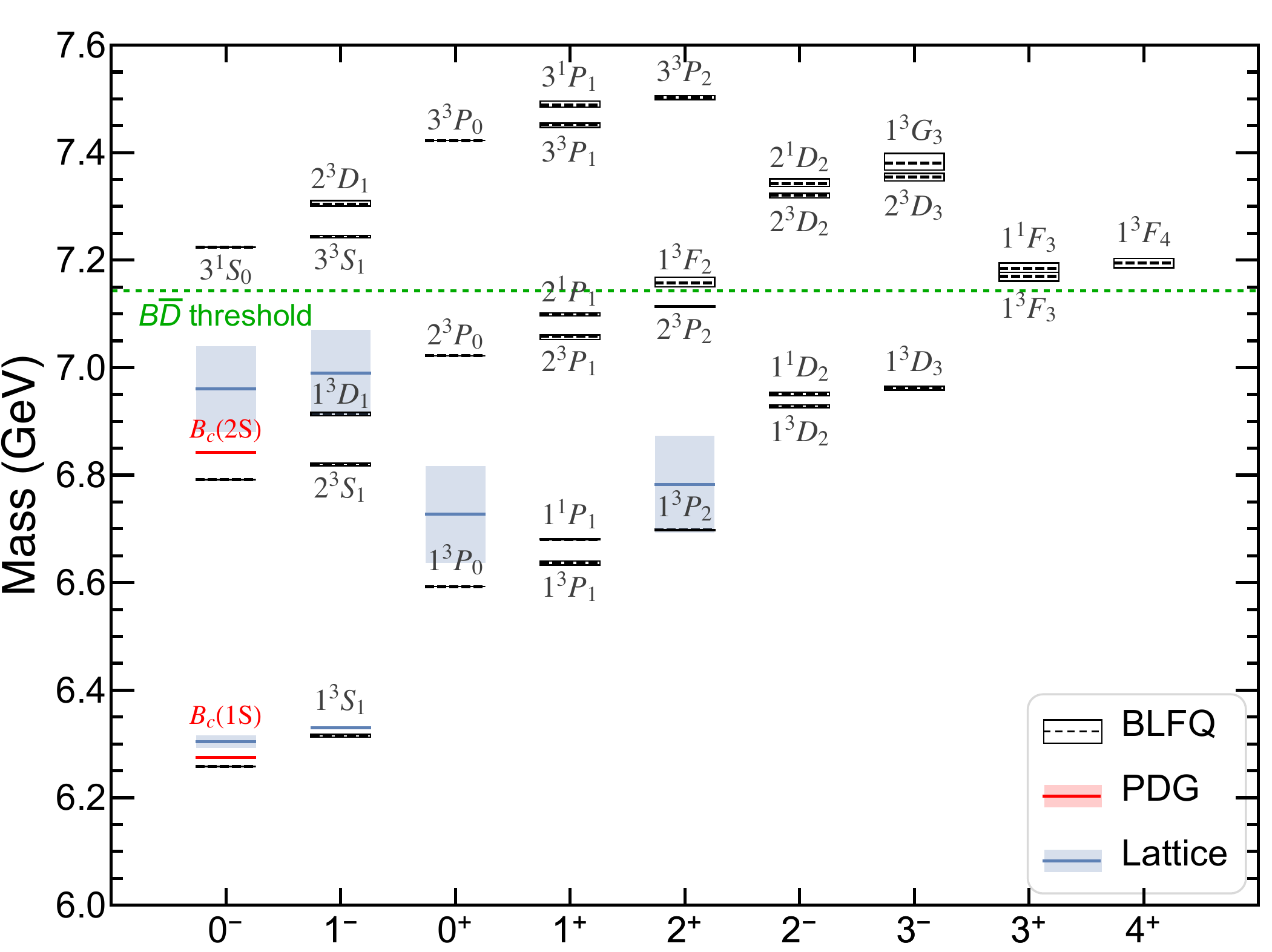}
		\caption{\label{fig1}The reconstructed $B_c\ (b\bar{c})$ spectrum at $N_\text{max}=L_\text{max}=32$. The horizontal axis is $J^{\mathsf{P}}$ and vertical axis is invariant mass in GeV. We compare with data from PDG \cite{1674-1137-40-10-100001} and Lattice  \cite{PhysRevLett.94.172001, PhysRevLett.104.022001, DAVIES1996131}, with central values shown as solid lines and uncertainties as shades.}
	\end{figure}

	In order to identify the multiplet of magnetic substates belonging to a single angular momentum $j$, the effective Hamiltonian is diagonalized for various $m_j$'s. One needs to perform the state identification to deduce the full set of quantum numbers $\mathsf{n}^{2s+1}\ell_j$ or $j^{\mathsf{P}}$, where $\ell$ is the orbital angular momentum, $\mathsf{n}$ is the radial quantum number (not to be confused with the basis quantum numbers $n$ and $l$). 
	The reconstructed mass spectrum up to the $B\overline{D}$ open flavor threshold  is presented in Fig. \ref{fig1}, where we use the dashed lines for the mean values of invariant masses:
	\begin{equation}
	\overline{M}\equiv \sqrt{\frac{M^2_{-j}+M^2_{1-j}+...+M^2_j}{2j+1}}.
	\end{equation}
	The boxes indicate the spread from different $m_j$, i.e. $\delta_jM\equiv \max (M_{m_j} )- \min( M_{m_j}) $, which is nonzero due to the violation of rotational symmetry arising from the Fock space and basis space truncations. 
	We also employ the mean spread to quantify the rotational symmetry violation from all high spin states below their respective dissociation thresholds,
	\begin{equation}
	\label{eq2}
	\overline{\delta_j M} \equiv \sqrt{\frac{1}{N_h} \sum_{h}^{j\neq0}(\delta_jM_h)^2} \quad \Big(N_h\equiv \sum_{h}^{j\neq0} 1\Big).
	\end{equation}
	We provide the experimental values \cite{1674-1137-40-10-100001} and results from Lattice \cite{PhysRevLett.94.172001, PhysRevLett.104.022001, DAVIES1996131} for comparison. 
	Note that the r.m.s deviations in TABLE \ref{tb1} are evaluated with respect to 8 and 14 states for charmonium and bottomonium \cite{PhysRevD.96.016022}, but only with respect to two experimental states for $B_c$.

\begin{figure}
		\centering
		\includegraphics[scale=0.45]{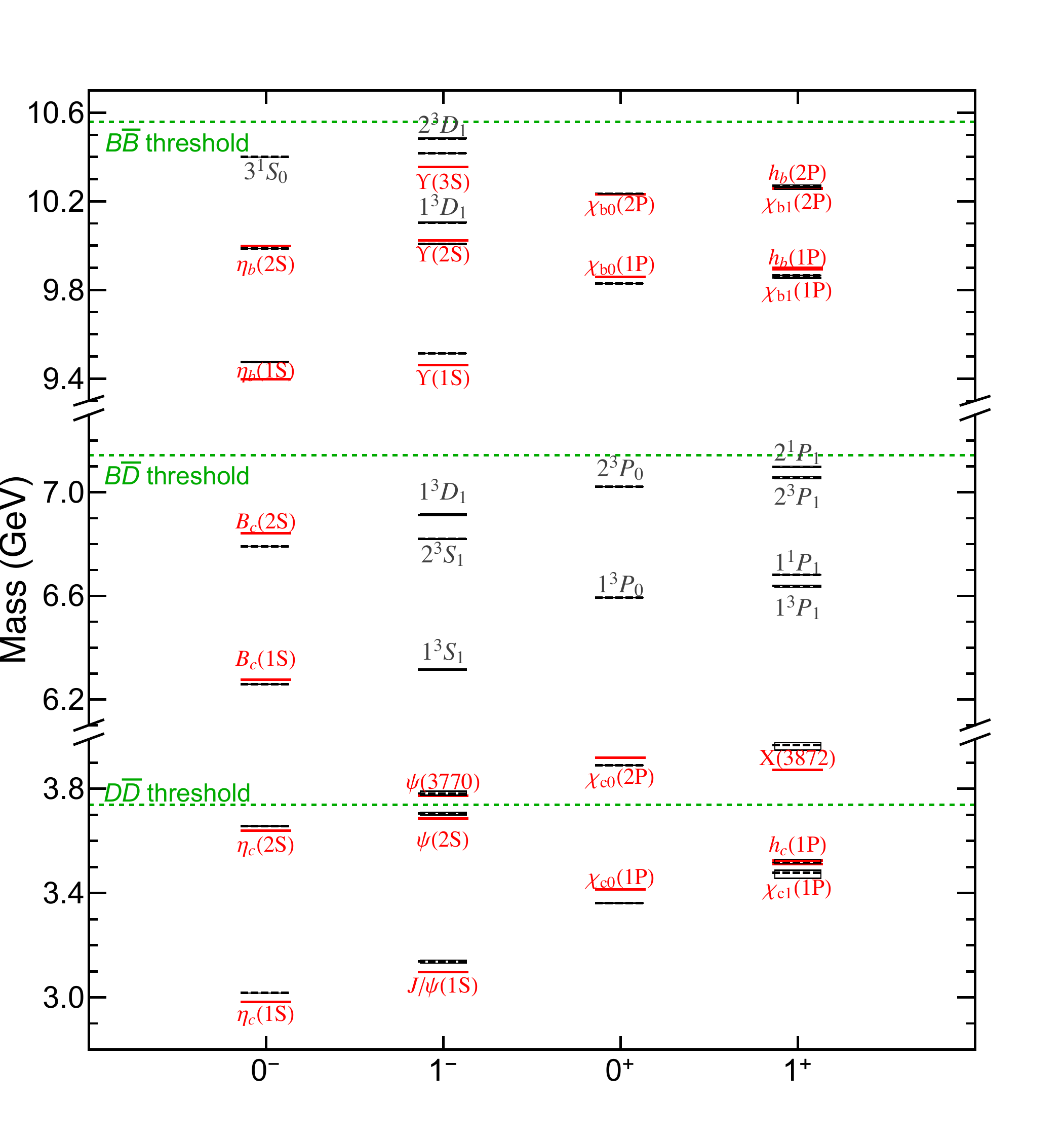}
		\caption{\label{fig7}The mass spectrum for charmonium, $B_c$, and bottomonium below their respective dissociation thresholds.  Note that, aside from an overall shift, the mass scales are similar. We only select the states with $J=0 \text{ and } 1$ for comparison. 
			Experimental results from PDG \cite{1674-1137-40-10-100001}  are in red while our results are in black.  The spread in $m_j$ values, defined in Eq. \ref{eq2}, is indicated by a rectangular black box around the $J = 1 $ theory results.  This is scarcely visible in many cases. 
			The mean spread of charmonium is larger than the other two. All the three systems have similar patterns in the spectrum while the heavier system has more states below the threshold.}
	\end{figure}

	We note that the mean spread of $b\bar{c}$, which is evaluated with a total of 12 states below the threshold of this system, is smallest in TABLE \ref{tb1}.  We compare the mass spectrum of $B_c$ and heavy quarkonia in Fig. \ref{fig7} for selected states. It is a challenge to visually ascertain which system exhibits the best rotational symmetry.  However, as evident from TABLE I, on the basis of the fraction of the mean spread relative to the total mass, we can see that the violation of rotational symmetry is larger for charmonium than for $B_c$ and bottomonium.  This suggests heavier systems retain rotational symmetry in our approach better than lighter systems.

	
	\subsection{Light-Front Wave Functions}
	Obtaining the light-front wave functions is a major motivation for this formalism, as they provide direct access to hadron observables. We present some of the valence LFWFs with different polarization and spin alignments for $B_c$ states. Specifically, we have the relation $m_j =s_1+s_2+m$, where $m$ is the orbital angular momentum projection.
	Since the phase $\exp(\imag m \theta)$ factorizes in the wave function on the two-body level, we drop it while retaining the sign for negative $\vec{k}_\perp$, i.e. we visualize the LFWFs at $k_y = 0$ ($\theta = 0$ and $\theta = \pi$). 
	
\begin{figure}
		\centering
		\includegraphics[scale=0.42]{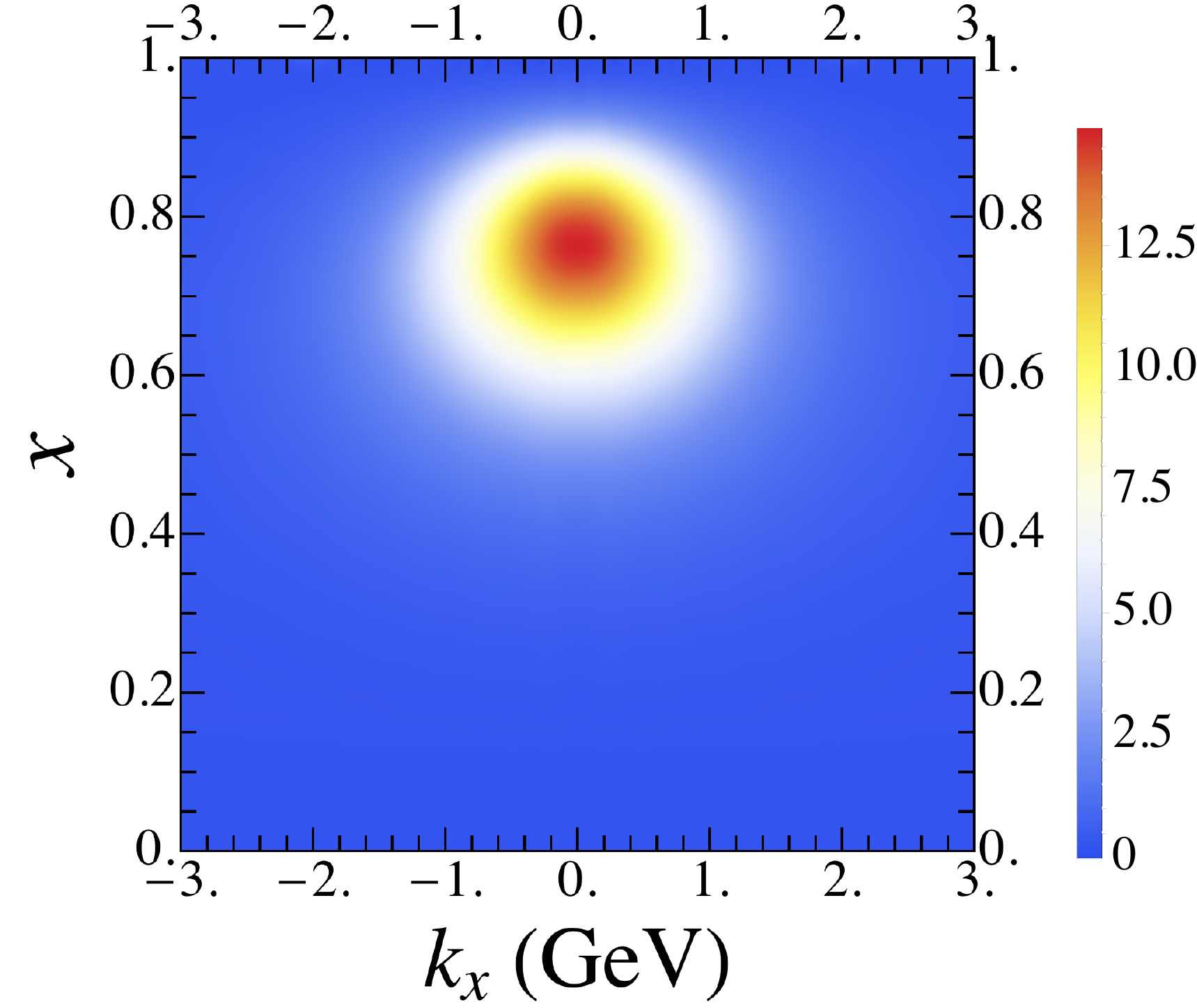} \hspace{1cm}
		\includegraphics[scale=0.42]{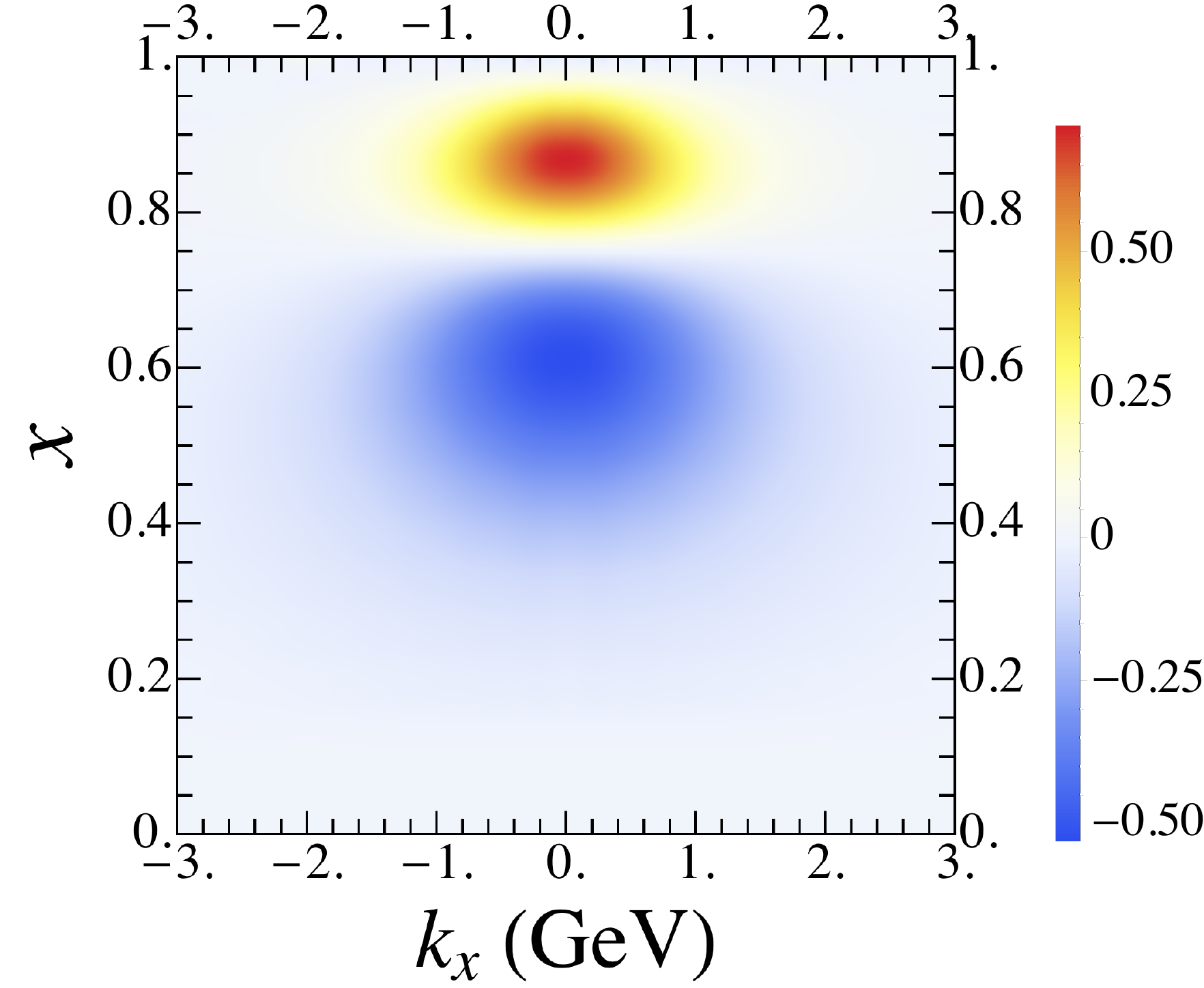}
		$$\text{a)}\ \psi_{\uparrow\downarrow - \downarrow\uparrow}(k_x, k_y = 0,x).  \text{\bf{ Left}}: (m+l)=\text{Even}; \ \text{\bf{Right}}: (m+l)=\text{Odd}. $$
		\includegraphics[scale=0.42]{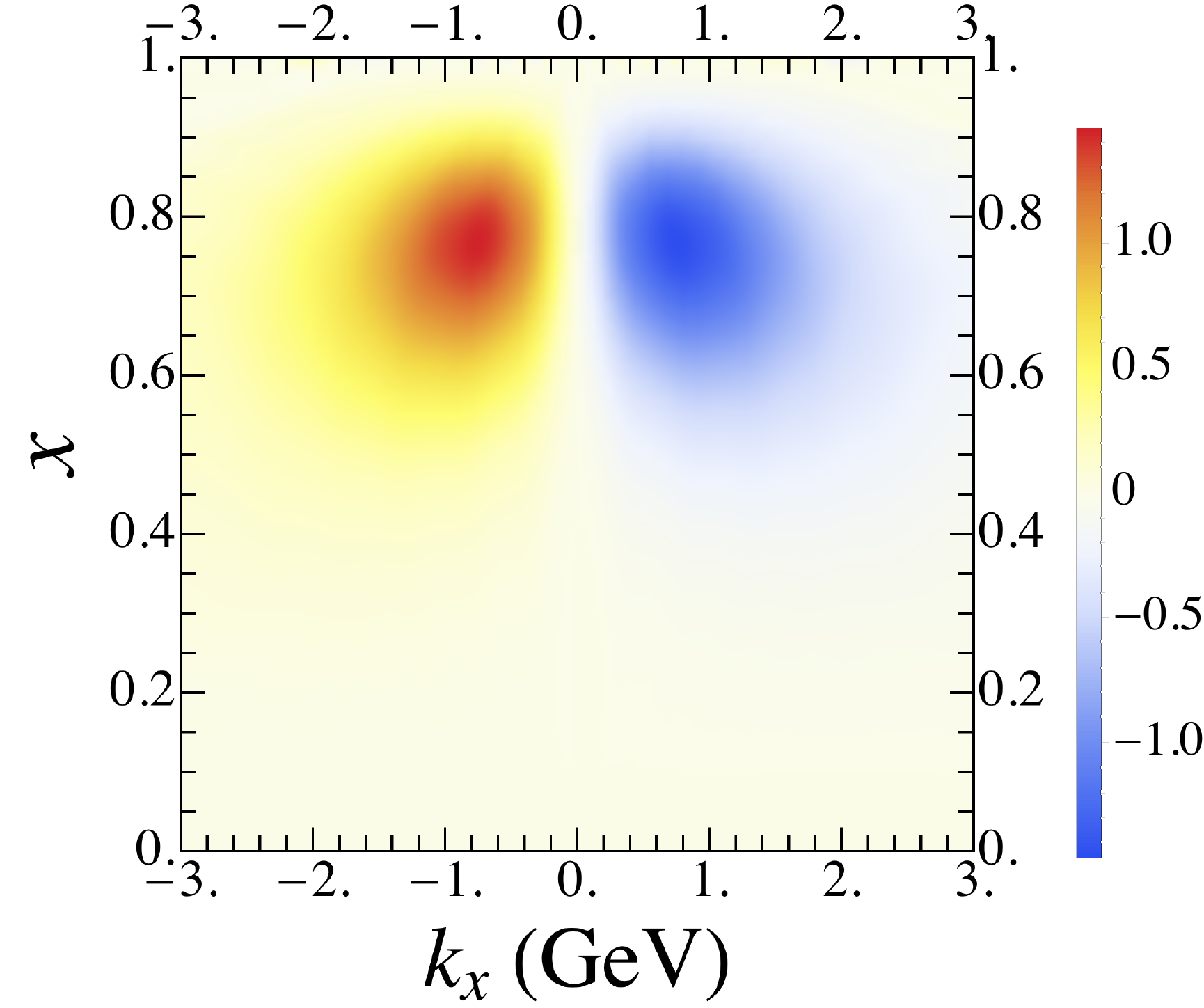} \hspace{1cm}
		\includegraphics[scale=0.42]{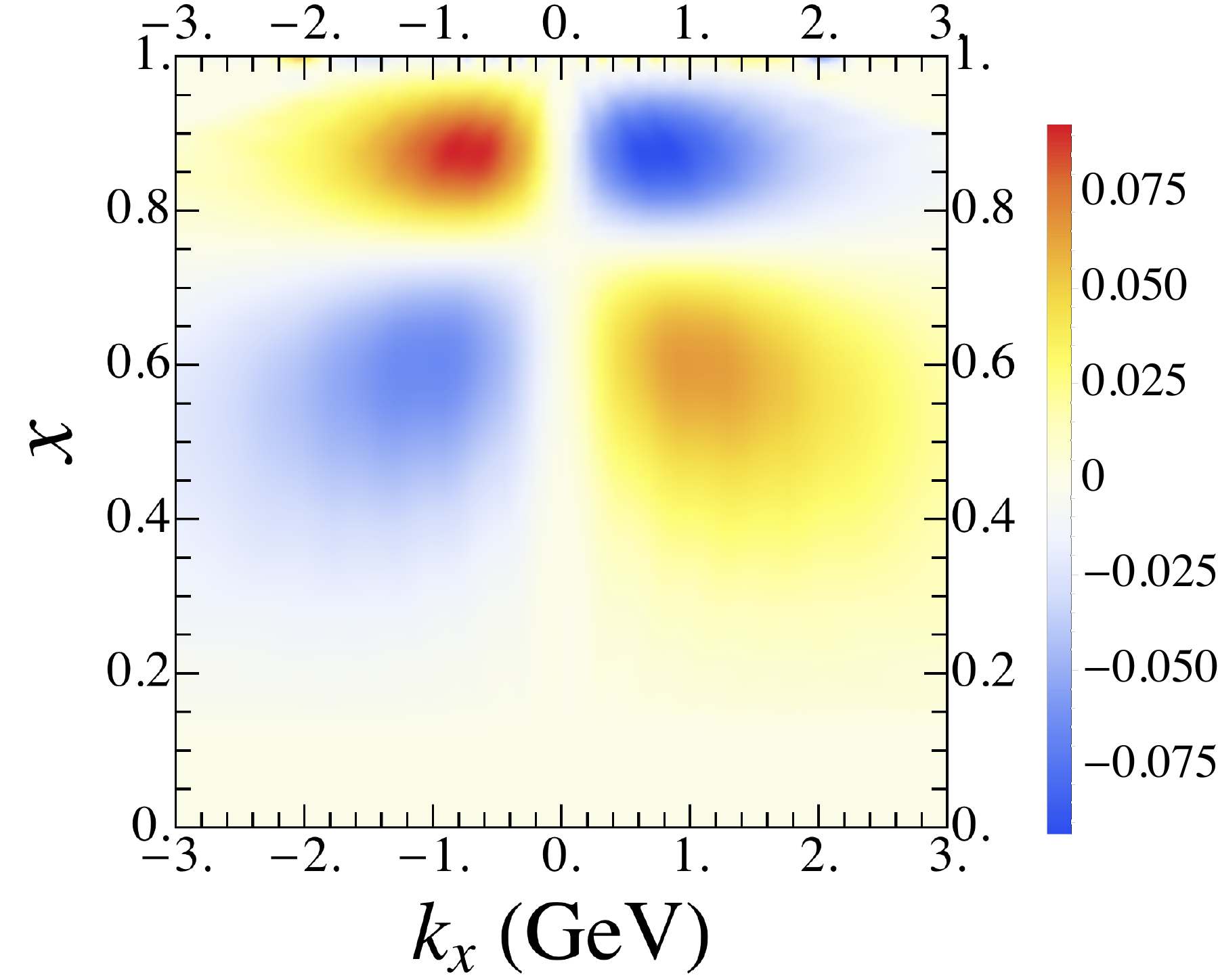}
		$$\text{b)}\ \psi_{\downarrow\downarrow}(k_x, k_y = 0,x)=\psi^*_{\uparrow\uparrow}(k_x, k_y = 0,x).  \text{\bf{ Left}}: (m+l)=\text{Even}; \ \text{\bf{Right}}:  (m+l)=\text{Odd}. $$
		\caption{\label{fig3}LFWFs of the ground state $B_c$ shown as plots of their magnitudes versus $x$ and $k_x$ at $k_y = 0$. In general spin alignment a) is dominant and reminiscent of non-relativistic behavior, while b) is purely a relativistic component. 
		}
	\end{figure}

	In Fig. \ref{fig3}, we show the ground state pseudoscalar LFWFs. There are two independent components with different spin alignments for $0^-$ state: $\psi_{\uparrow\downarrow - \downarrow\uparrow}(\vec{k}_\bot,x) \equiv \frac{1}{\sqrt{2}}[\psi_{\uparrow\downarrow}(\vec{k}_\bot,x) - \psi_{\downarrow\uparrow}(\vec{k}_\bot,x)] $ and $\psi_{\downarrow\downarrow}(\vec{k}_\bot,x)=\psi^*_{\uparrow\uparrow}(\vec{k}_\bot,x)$.
	The former is dominant and reduces to the non-relativistic wave function in the heavy quark limit, while the latter is of pure relativistic origin. 
	
	
	Furthermore, $B_c$ has another significant feature that distinguish from the quarkonium (equal-mass mesons).
	For the heavy quarkonia, charge conjugation is a good symmetry, and is reflected by states having components either even or odd in $(m+l)$ \cite{LI2016118}. There is no charge conjugation symmetry of $B_c$, but we do observe that our solutions are dominated by either even or odd $(m+l)$. TABLE \ref{tb4} exhibits this dominance, along with the comparison with heavy quarkonia of the ground states.  In a separate test calculation, we verified that, as the mass difference between quark and anti-quark decreases, the contribution from even $(m+l)$  is getting smaller, and progresses smoothly to the equal-mass limit.

	\begin{table}
		\centering 
		\begin{tabular}{c|c c|c c}
			\hline\hline
			\\[-0.3cm]
			\multirow{2}{*}{ \diagbox[width=3.5cm, height= 1.1cm]{system}{\raisebox{2cm}{\rotatebox{0}{ \footnotesize{even/odd $(m+l)$}}}}	} &  \multicolumn{2}{c}{  \hspace{1cm} $\abs{\psi_{\uparrow\downarrow - \downarrow\uparrow}(\vec{k}_\bot,x)}^2$}  \hspace{1cm}&    
			\multicolumn{2}{|c}{ $\quad \abs{\psi_{\downarrow\downarrow}(\vec{k}_\bot,x)}^2 +\abs{\psi_{\uparrow\uparrow}(\vec{k}_\bot,x)}^2$}    \\[0.2cm]
			
			&  \hspace{0.6cm} Odd & Even & \hspace{0.5cm}  Odd & Even \\[0.05cm]
			\hline
			$c\bar{c}$	& \hspace{0.6cm} $88.01\%$  		& $0$		& \hspace{0.5cm} $11.99\%$ 	&$0$		\\
			$b\bar{c}$	&  \hspace{0.6cm} $91.62\%$		&$0.35\%$		& \hspace{0.5cm} $7.98\%$ 		&$0.05\%$	\\
			$b\bar{b}$	& \hspace{0.6cm} $96.61\%$		&$0$		& \hspace{0.5cm} $3.39\%$ 		&$0$	\\
			\hline\hline
		\end{tabular}	
		\caption{	\label{tb4} The probabilities of finding the specified even or odd $(m+l)$ in the  ground state of heavy mesons. The dominant spin alignment listed here are the components that persist in the non-relativistic limit.  Note the systematic increase of these dominant components with the increasing meson mass.}  
	\end{table}

	The LFWFs of $B_c$ do not have symmetry with respect to $x= \frac{1}{2}$ in the longitudinal direction, another feature distinguishing $B_c$ from heavy quarkonium. The wave function peaks in $x$ near the bottom quark mass fraction, i.e. $x = m_b/(m_b+m_{\bar{c}}) \approx 0.75$, as expected from the major role of the kinetic energy terms in the Hamiltonian.  This asymmetry will have interesting consequences for other observables as discussed 
	in the following subsections.

	\subsection{Decay Constants}
	
	Meson decay constants, $f_h$, are hadronic properties defined from the matrix element of the local current that annihilates the meson. They are
	\begin{equation}
	\begin{gathered}
	\matrixel{0}{\bar{c}\gamma^\mu\gamma_5 b }{P(p)}= \imag p^\mu f_P,\\
	\matrixel{0}{\bar{c}\gamma^\mu b}{V(p,\lambda)}= e^\mu_\lambda M_Vf_V,
	\end{gathered}
	\end{equation}
	for pseudoscalar ($P$) and vector ($V$) states, respectively. Here $p$ is the momentum of the meson, and $e^\mu_\lambda$ is the polarization vector:
	\begin{equation}
	e^\mu_\lambda(k) = \big(e_\lambda^-(k), e_\lambda^+(k), \vec{e}_{\bot\lambda}(k)\big) \triangleq 
	\left \{
	\begin{aligned}
	&\Big(\frac{\vec{k}^2_\bot - M_V^2}{M_V k^+} , \frac{k^+}{M_V} , \frac{\vec{k }_\bot}{M_V}\Big),\  \lambda=0\\
	&\Big(\frac{2\vec{\epsilon} _{\bot \lambda} \cdot \vec{k }_\bot}{k^+} ,0 , \vec{\epsilon}_{\bot\lambda}\Big), \quad  \lambda=\pm 1\\
	\end{aligned}
	\right. ,
	\end{equation}  
	where $ \vec{\epsilon}_{\bot \pm} = (1,\pm \imag)/\sqrt{2}$, and we adopt $\lambda \equiv  m_j$ as the angular momentum projection.
	The decay constant can be computed in the light-front representation in terms of LFWFs with different polarizations and corresponding current components.
	In this work, we choose the ``good current'' ($\mu = +$) and the longitudinal polarization $(\lambda = 0)$ for the calculations. Since for $J=0$ states, $+$ and $\bot$ currents lead to identical results; for $J=1$ states, it has been illustrated that $\lambda=0$ and $\lambda=1$ provide comparable results for S-waves \cite{PhysRevD.98.034024}. 
	This choice leads to the decay constant as:
	\begin{equation}
	\frac{f_{P,V}}{2\sqrt{2 N_c}}= \int_0^1 \frac{\dd x}{2\sqrt{x(1-x)}} \int \frac{\dd ^2 k_\bot}{(2\pi)^3} \psi^{(\lambda=0)}_{\uparrow\downarrow \mp \downarrow\uparrow} (\vec{k}_\bot,x),
	\end{equation}
	where the ``minus'' and ``plus'' signs correspond to pseudoscalar and vector states, respectively.
	Here, calculations have been done with $N_\text{max} = 32$, corresponding to $\Lambda_\text{UV}\triangleq \kappa \sqrt{N_\text{max}} \approx m_b+m_{\bar{c}}$, where $\Lambda_\text{UV}$ is the ultraviolet regulator. This is to balance the needs for better basis resolution and lower UV scale owing to the omitted radiative corrections. 
	An early effort using QCD sum rules provided $300$ MeV as an estimate for $f_{B_c}$ and $500$ MeV\footnote{T. Aliev, private communication.} for $f_{B_c^*}$ \cite{Aliev1992}. We present a survey of recent work in TABLE \ref{tb2} along with results from Lattice \cite{PhysRevD.96.054501,PhysRevD.91.114509} and other approaches (see Refs. \cite{Baker2014,Wang2013,PhysRevD.80.054016,PhysRevD.97.054014,PhysRevD.81.034010,PhysRevD.59.094001}) for comparison. Lattice results are systematically smaller than the other methods by about $20\%$, as also discussed in other references (e.g. \cite{BENBRIK2009172,Baker2014}). 
	
\begin{figure}
		\centering
		\includegraphics[scale=0.5]{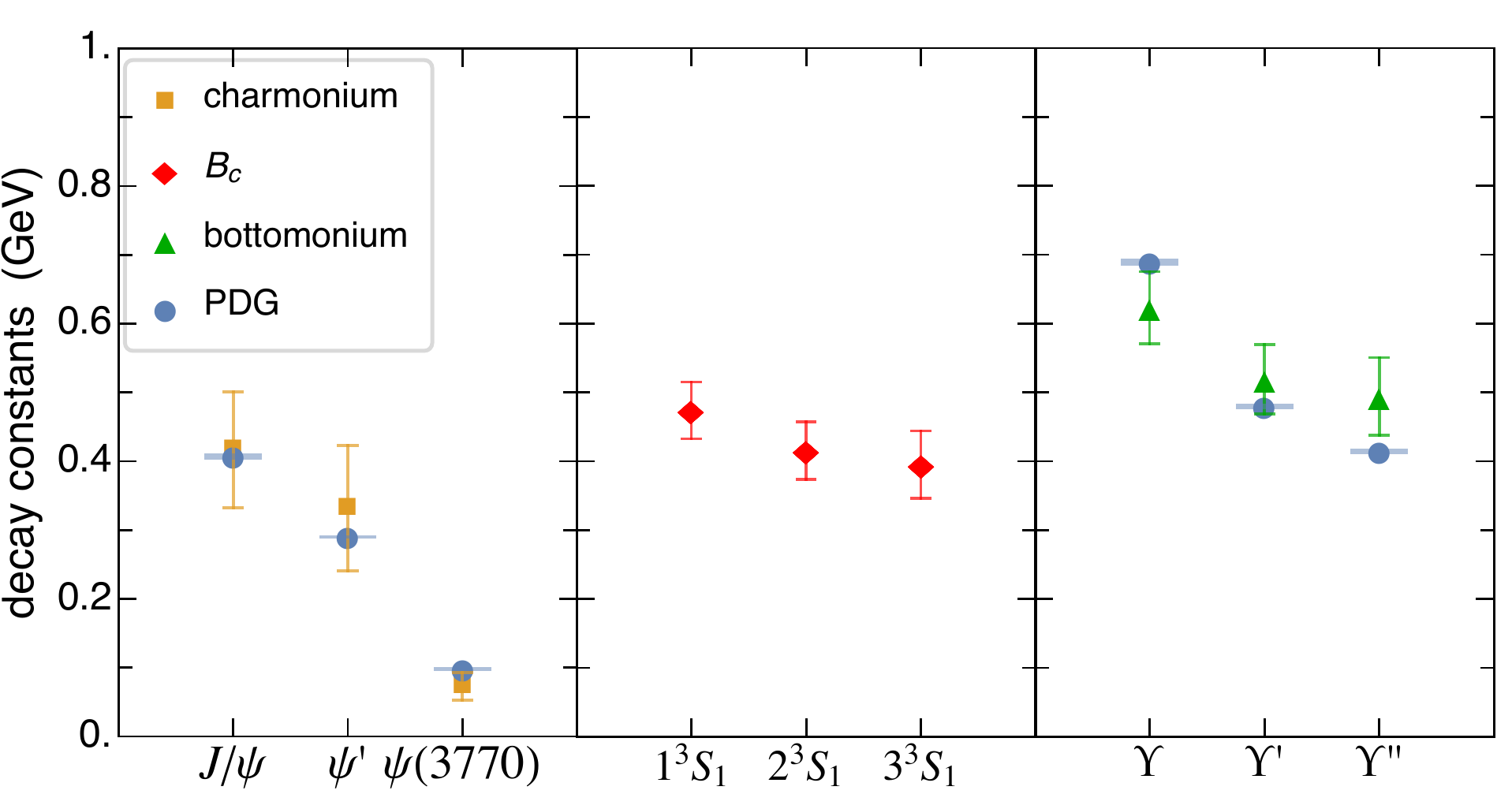}
		\caption{\label{fig2}The decay constants for vector states of charmonium, $B_c$ and bottomonium. Results of charmonium and bottomonium are from previous work \cite{PhysRevD.96.016022} and PDG \cite{1674-1137-40-10-100001}.}
	\end{figure}

	In Fig. \ref{fig2}, we compare the vector decay constants of $B_c$ and heavy quarkonia \cite{PhysRevD.96.016022}. We can see a trend that decay constants within each meson system decrease with increasing radial quantum numbers.
	This trend seems reasonable since increasing radial quantum numbers correspond to less binding and a larger spread in the radial probability distributions.
	In addition, we note that  the vector decay constants increase with the mass of the system for corresponding states, e.g. $J/\Psi < B_c (1^3S_1 )< \Upsilon$. This trend correlates with decreasing size as the mass increases which is discussed further in the next session.
	
	\begin{table}
		\centering 
		\begin{tabular}{ccccccccccccc}
			\hline\hline
			Constant (MeV)  &  \hspace{0.1cm }this work   \hspace{0.1cm }&  \hspace{0.1cm } Lattice \cite{PhysRevD.96.054501,PhysRevD.91.114509}   \hspace{0.1cm } &  \hspace{0.1cm } QCD sum rules \cite{Baker2014,Wang2013}   \hspace{0.1cm }     &  \hspace{0.1cm } LFQM \cite{PhysRevD.80.054016}    \hspace{0.1cm }&   CCQM \cite{PhysRevD.97.054014}  &  \hspace{0.1cm } BSE  \cite{PhysRevD.59.094001}  \\
			\hline
			$f_{B_c}$		&$523(62)$  		& $427(6)$		&$528(19)$    & $551$		&$489.3$ 		&$578$	\\
			$f_{B^*_c}$		&$474(42)$		&$422(13)$		& $384(32)$ 		& $508$ 	& 	&	\\
			\hline\hline
		\end{tabular}
		\caption{Pseudoscalar and vector decay constants of the ground state $B_c$ and its vector partner $B^*_c$. The uncertainties of this work indicate the sensitivity to basis truncation, which is taken to be $\Delta f_{b\bar{c}} =2 \abs{f_{b\bar{c}}(N_\text{max}=32) - f_{b\bar{c}}(N_\text{max}=24)}$.}  
		\label{tb2}
	\end{table}
	
	\subsection{Charge and Longitudinal Momentum  Densities}
	
\begin{figure}
		\centering
		\hspace{-2cm}
		\includegraphics[scale=0.5]{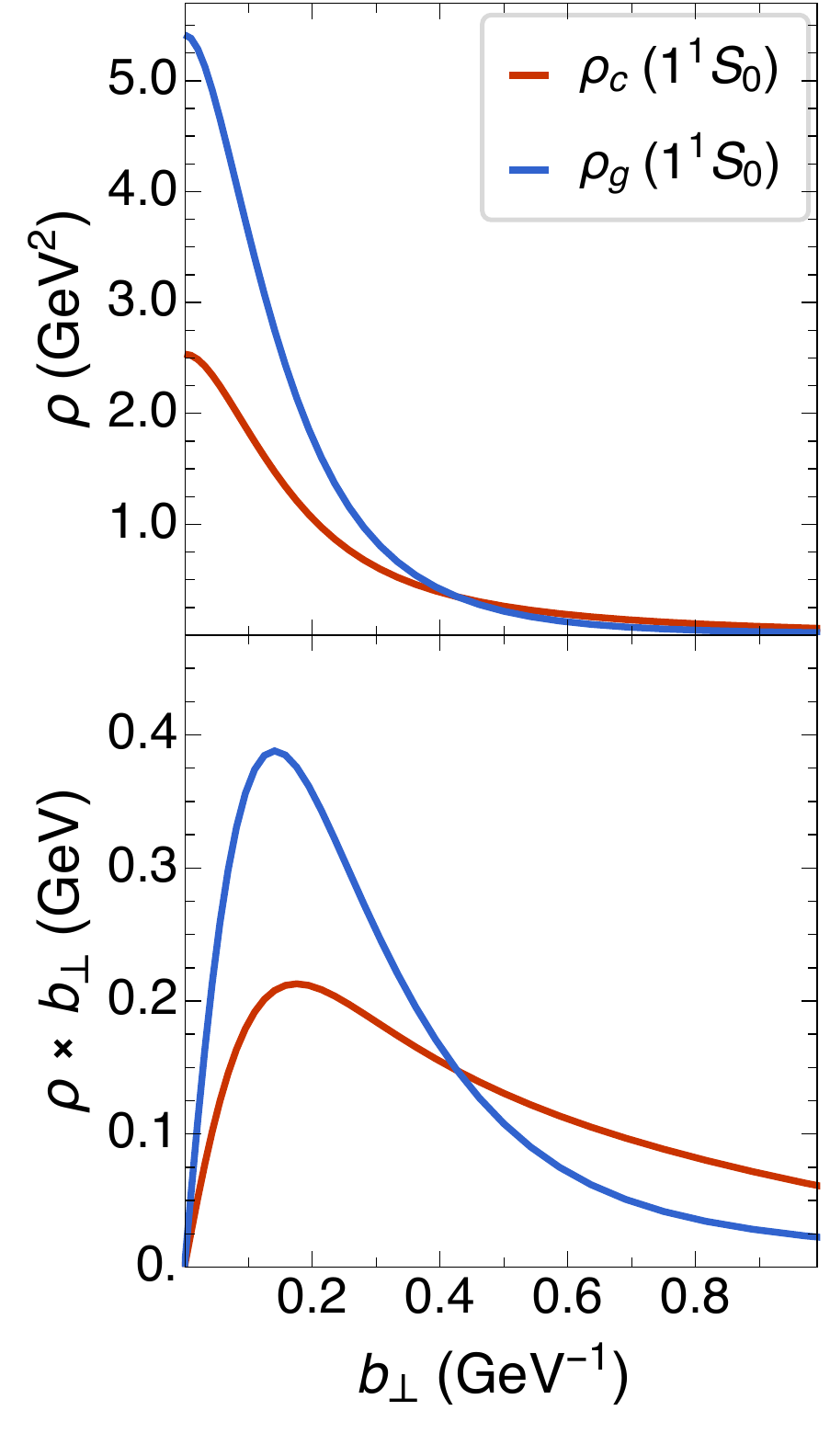} \includegraphics[scale=0.5]{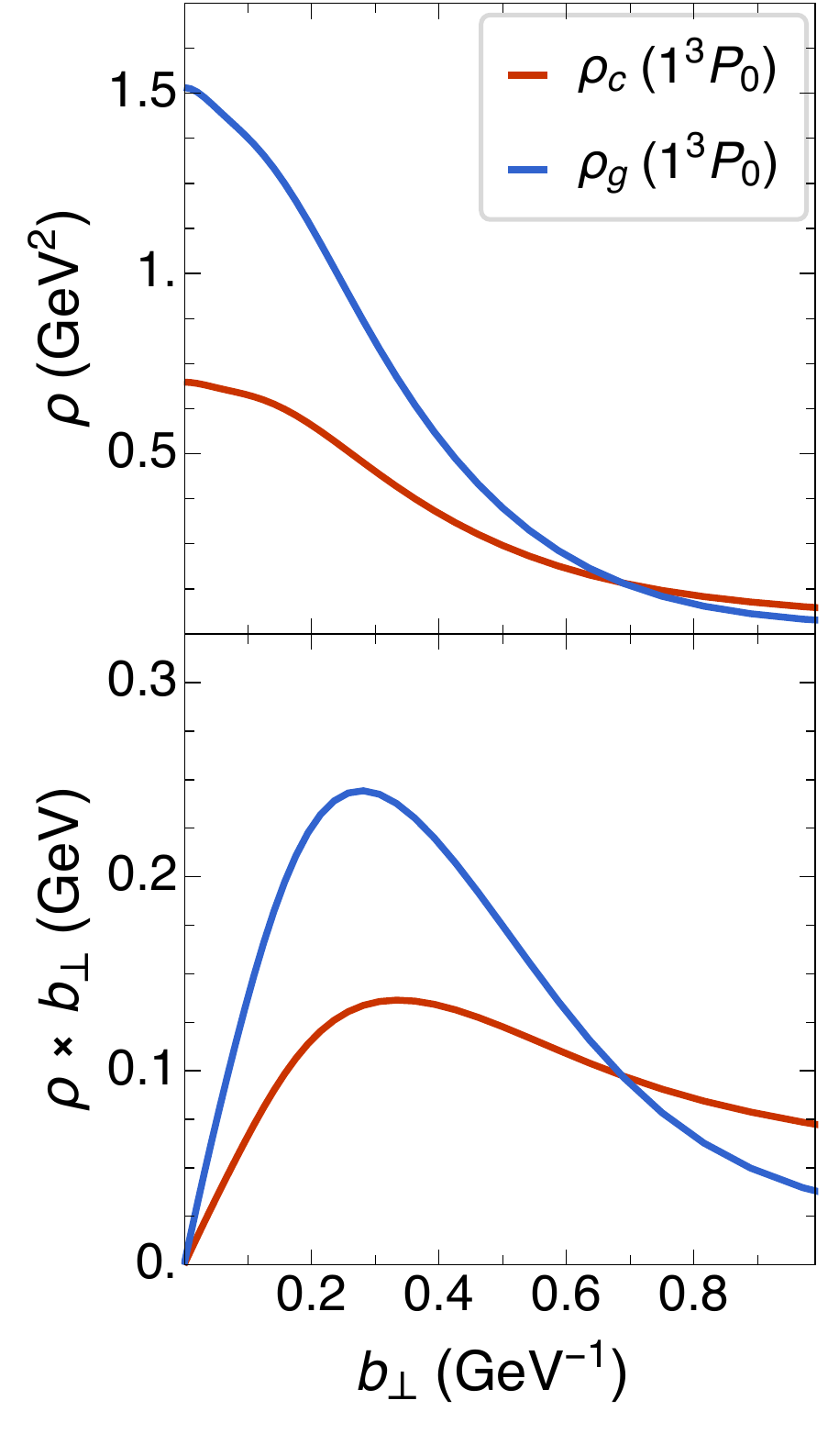} \includegraphics[scale=0.5]{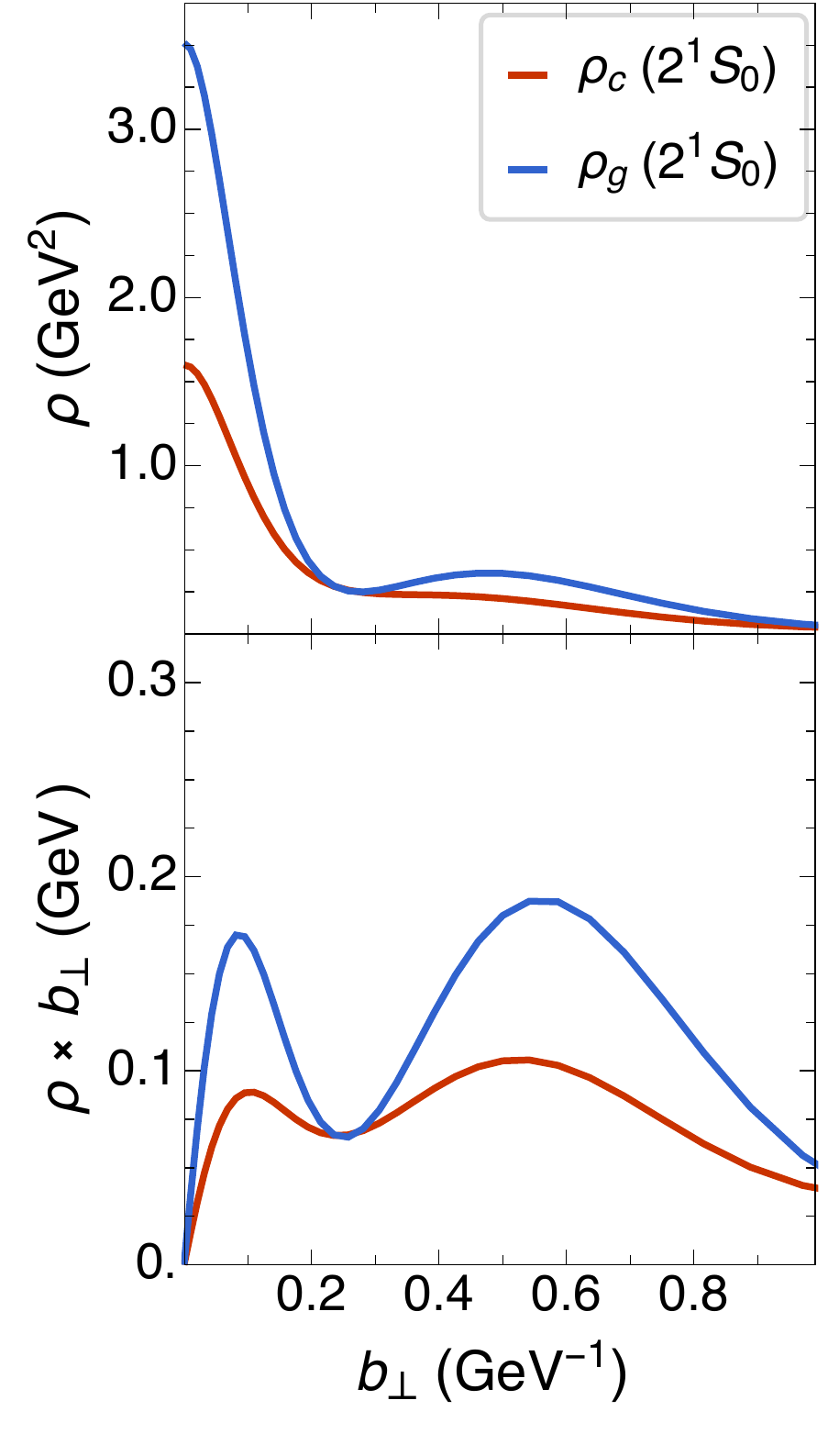}
		\hspace{-2cm}
		
		\caption{\label{fig6}The charge density and longitudinal momentum density on the transverse plane of the pseudoscalar and scalar states.}
	\end{figure}

	The transverse density offers insight into the hadron structure. In this work, we study the charge density in the transverse impact parameter space of $B_c$ mesons. By definition, it is the two-dimensional Fourier transform of the Dirac form factor \cite{burkardt2003impact,PhysRevLett.99.112001},
	\begin{equation}
	\rho_c(\vec{b}_\bot)= \int \frac{\dd ^2 \Delta_\bot}{(2\pi)^2} e^{\imag \vec{\Delta}_\bot \cdot \vec{b}_\bot} F_1 (q^2 = -\vec{\Delta}^2_\bot),
	\end{equation}
	where $\vec{\Delta}_\bot$ is the transverse momentum transfer, and $\vec{b}_\bot$ can be interpreted as the conjugated position of $\vec{\Delta}_\bot$ at which the current probes the charge density. Analogous to the charge distribution, we can perform the two-dimensional Fourier transform of the gravitational form factor, which can be interpreted as the longitudinal momentum density in the transverse plane \cite{PhysRevD.78.071502}.
	In the LFWF representation of the two-body ($b\bar{c}$) approximation, they can be expressed as,
	\begin{equation}
	\rho_c(\vec{b}_\bot) = \frac{1}{3}  \sum_{s,\bar{s}} \int_0^1 \frac{\dd x}{4\pi (1-x)^2} \abs{\widetilde{\psi}_{s\bar{s}} \big( \frac{-\vec{b}_\bot}{1-x},x \big)}^2  +
	\frac{2}{3}  \sum_{s,\bar{s}}\int_0^1 \frac{\dd x}{4\pi x^2} \abs{\widetilde{\psi}_{s\bar{s}} \big( \frac{\vec{b}_\bot}{x},x \big)}^2  ,
	\end{equation}
	\begin{equation}
	\rho_g(\vec{b}_\bot) =  \sum_{s,\bar{s}} \int_0^1 \frac{\dd x}{4\pi}\frac{x}{(1-x)^2} \abs{\widetilde{\psi}_{s\bar{s}} \big( \frac{-\vec{b}_\bot}{1-x},x \big)}^2  +
	\sum_{s,\bar{s}}\int_0^1 \frac{\dd x}{4\pi} \frac{1-x}{x^2} \abs{\widetilde{\psi}_{s\bar{s}} \big( \frac{\vec{b}_\bot}{x},x \big)}^2  .
	\end{equation}
	Each density is normalized to unity, as the unit charge and the mass of the meson, respectively. The momentum density is more concentrated in the center than the charge density, where the difference is a relativistic effect \cite{PhysRevD.96.016022}. This pattern can be observed in Fig. \ref{fig6}, where we present the results  of pseudoscalar and scalar states. We compare the r.m.s radii of $\rho_g(\vec{b}_\bot)$ among heavy meson systems, which are $0.84$ GeV$^{-1}$, $0.58$ GeV$^{-1}$, $0.57$ GeV$^{-1}$ for $J/\Psi$,  $B_c (1^3S_1 )$, and $ \Upsilon$, respectively. This result is consistent with the trend of decay constants: for the heavier system, it has smaller the radii, therefore it is easier to decay.
	
	%
	
	\subsection{Distribution Amplitude}
	
\begin{figure}
		\centering
		\includegraphics[scale=0.5]{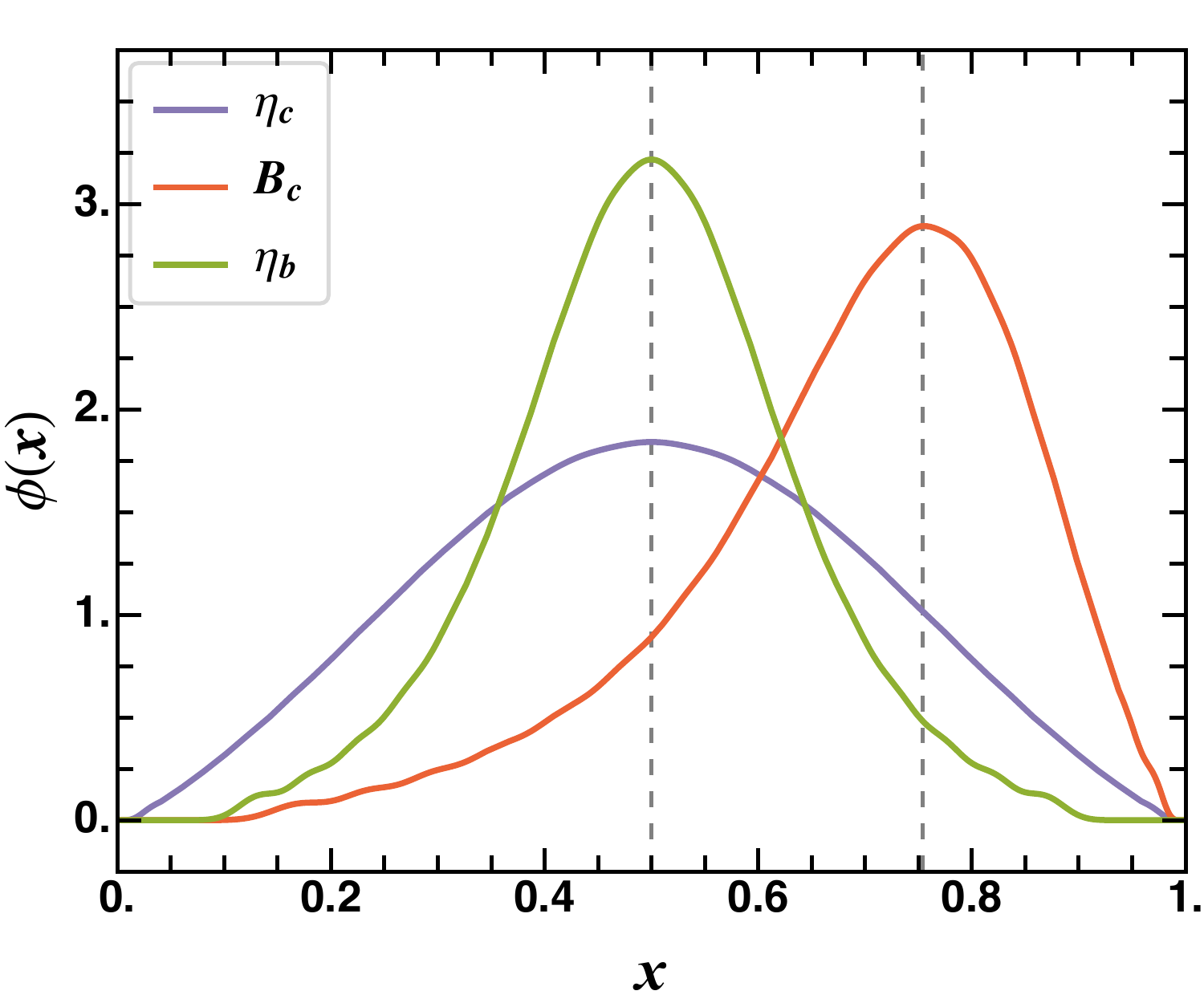}
		\hspace{1cm}
		\includegraphics[scale=0.5]{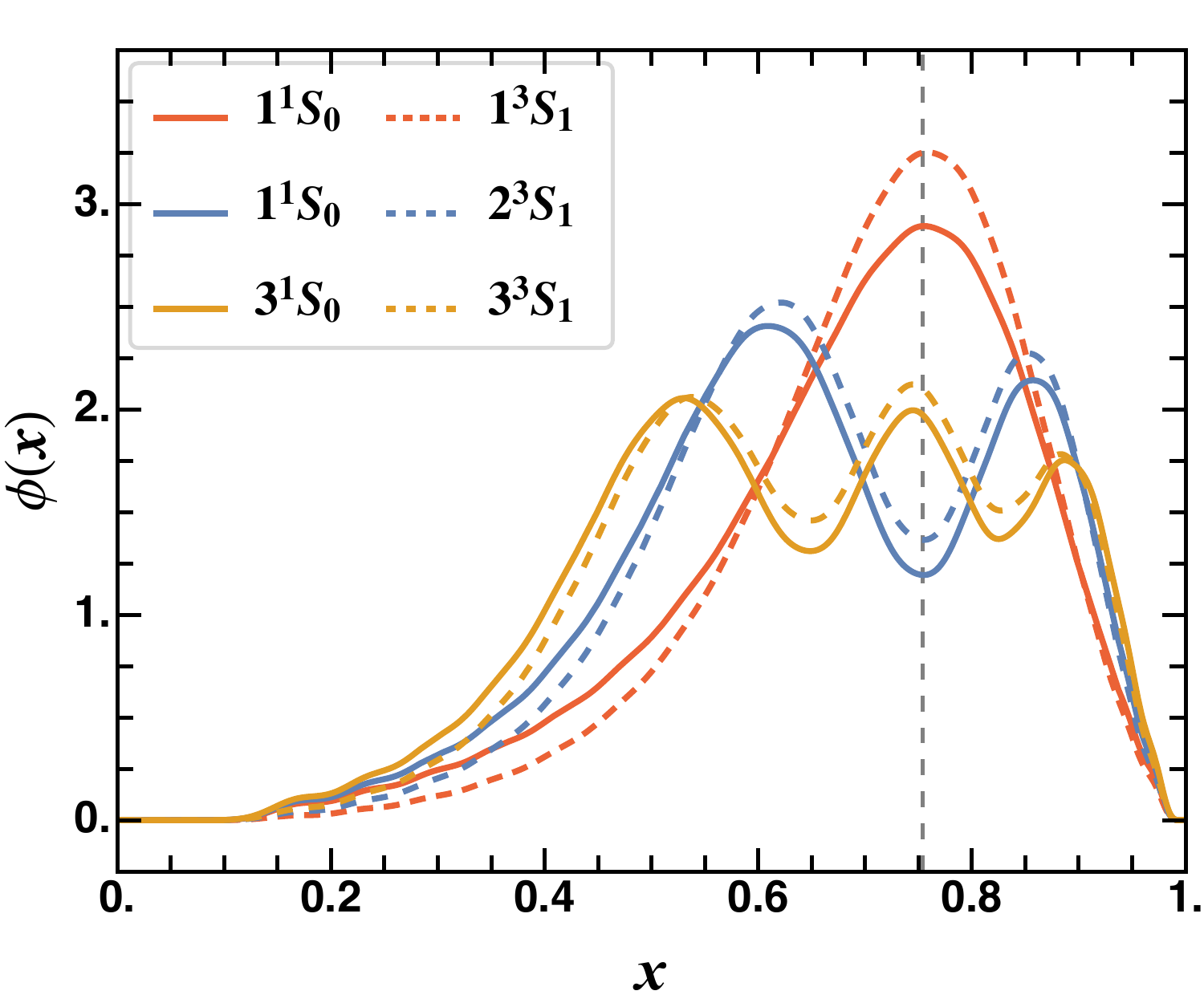}
		\caption{\label{fig8}The distribution amplitudes(DAs) of the ground states of charmonium, $B_c$, and bottomonium (left panel). DAs of the pseudoscalar and vector $B_c$'s and their radial excitations (right panel). 
			Vertical dashed lines indicate the constituent quark mass fraction of the corresponding system. Specifically peaks of the DAs of quarkonium occur at $x=m_c/(m_c+m_{\bar{c}})=m_b/(m_b+m_{\bar{b}})=1/2$, and the peak of $B _c$ is close to $x = m_b/(m_b+m_{\bar{c}})\approx0.75$.}
	\end{figure}

	Distribution amplitudes (DAs) are defined from the light-like vacuum-to-meson matrix elements, and can be written with LFWFs as,
	\begin{equation}
	\frac{f_{P,V}}{2\sqrt{2N_c}}\phi_{P,V}(x) = \frac{1}{\sqrt{x(1-x)}} \int\frac{\dd ^2 \vec{k}_\bot}{2(2\pi)^3} \psi^{\lambda=0}_{\uparrow\downarrow\mp \downarrow\uparrow} (\vec{k}_\bot,x)
	\end{equation} 
	with $f_{P,V}$ the decay constants for pseudoscalars and vectors, respectively. Note that DAs defined here are normalized to unity. We compare the ground states pseudoscalar DAs of the charmonium, $B_c$ meson, and bottomonium in Fig. \ref{fig8}. The width of the DA decreases while the peak height increases with the mass of the system, and approaches a $\delta$-function in the non-relativistic limit. For charmonium and bottomonium, peaks are at $x=1/2$ due to the equal mass of the constituent quark and anti-quark. While for $B_c$, the peak is close to the constituent quark mass fraction, i.e.  $x= m_b/(m_b+m_{\bar{c}}) \approx 0.75$, which is consistent with the distribution of the LFWFs in the previous section. We present the ground-state pseudoscalar and vector DAs and their excited states. Note that the  pseudoscalar and vector DAs have similar patterns but they are not identical. This is due to the different configuration mixings as controlled by the one-gluon exchange interaction in this model. The radial excited states have important distinctions: dips appear with the radial excitations.
	This pattern also appears in charmonium and bottomonium in BLFQ and in other methods \cite{PhysRevD.77.034026,Hwang2009}. Wiggles near both extremes of $x$ arise from the limited range of basis spaces employed. 
	
	\section{\label{sec4}Summary}
	
	In this work, we investigated the unequal-mass meson system $B_c \ (b\bar{c})$ with the BLFQ approach. All model parameters are fixed by reference to charmonium and bottomonium systems.  We found reasonable agreement with existing experiments and with other theoretical calculations. 
	
	We carried out the calculations with the basis limit $N_\text{max} =L_\text{max} = 32$, which corresponds to the specific UV (IR) regulator $b\sqrt{N_\text{max}} \approx 6.77 \text{ GeV}\ (b/\sqrt{N_\text{max}} \simeq 0.21 \text{ GeV})$. We first predicted the $B_c$ mass spectrum
	and presented the LFWFs of some selected states. These results are obtained from diagonalizing a light-front Hamiltonian based on light-front holography. We discussed a significant difference  between the LFWFs of $B_c$  and heavy quarkonium: only the unequal-mass system $B_c$ allows both positive and negative charge parities in the wave functions, due to the absence of charge conjugation symmetry.
	
	We calculated other observables with LFWFs such as the decay constants of pseudoscalar and vector states.
	As additional applications and tests of our model, we calculated transverse charge and momentum densities. 
	
	Our successes here in applications of the model for heavy quarkonium to the unequal mass heavy meson system provide support for further extensions to lighter systems.  One anticipates greater challenges to the model which are likely to require the inclusion of a dynamical gluon in a higher Fock sector. This naturally incorporates self-energy processes and raises challenging issues of renormalization \cite{PhysRevD.77.085028,PhysRevD.92.065005}.  Additional significant physics should also be included such as chiral symmetry breaking, etc.
	
	\section{Acknowledgments}
	
	We wish to thank Shaoyang Jia, Meijian Li, Wenyang Qian and Anji Yu for valuable discussions. We also thank Dr. Zhigang Wang for clarifying the result from the QCD sum rule \cite{Wang2013} that we quote in TABLE \ref{tb2}. 
	P. Maris thanks the Fundação de Amparo \`{a}  Pesquisa do Estado de São Paulo (FAPESP) for support under grand No. 2017/19371-0.
	This work was supported in part by the Department of Energy under Grant No. DE-FG02- 87ER40371, DE-SC0018223 (SciDAC-4/NUCLEI), DE-SC0015376 (DOE Topical Collaboration in Nuclear Theory for Double-Beta Decay and Fundamental Symmetries) and DE-FG02-04ER41302. Computational resources were provided by the National Energy Research Supercomputer Center (NERSC), which is supported by the Office of Science of the U.S. Department of Energy under Contract No. DE-AC02-05CH11231.
	
	\bibliography{bc}

\end{document}